\documentclass[
  draft            
  ]
  {aipproc}

\usepackage{epsfig}
\layoutstyle{6x9}

\begin{document}

\title{NUFACT'03: The Fate of the Clones}

\author{A. Donini\footnote{E-mail andrea.donini@roma1.infn.it}}{
 address={IFT, Universidad Autonoma Madrid, Cantoblanco 28049 Madrid, Spain}
}

\begin{abstract}
We present a Neutrino-Factory-based setup with three detectors of different kind
in principle capable to solve the eightfold-degeneracy in the simultaneous measurement
of $\theta_{13}$ and $\delta$, for $\theta_{13} \geq 1^\circ$ 
($\sin^2 (2 \theta_{13}) \geq 10^{-3}$). Our setup includes a Superbeam-driven 
water Cherenkov (the Superbeam conceived as the first stage of the Neutrino Factory); 
two muon-storage-ring-driven detectors (namely, a large magnetized iron calorimeter and 
an emulsion cloud chamber) to take advantage of both the so-called
``golden'' ($\nu_e \to \nu_\mu$) and ``silver'' ($\nu_e \to \nu_\tau$) channels.
\end{abstract}

\maketitle

The planned long baseline experiments~\cite{planned}
will improve the measurement of $\Delta m^2_{atm}$ and of $\theta_{23}$
and measure or increase the bound on $\theta_{13}$~\cite{komatsu,minosIN} 
(see also~\cite{Migliozzi:2003pw}). This new generation of experiments,
however, is only the first step of a long-lasting experimental program 
including the development of some ``superbeam'' facilities (whose combination 
can strongly improve our knowledge on $\theta_{13}$, see~\cite{Huber:2002mx}) 
and, eventually, of a ``Neutrino Factory'' \cite{Geer:1998iz,DeRujula:1998hd}. 
One of the main goals of the Neutrino Factory program (see for example 
\cite{Blondel:2000gj,Apollonio:2002en} and refs. therein) would be 
the discovery of leptonic CP violation and, possibly, 
its study~\cite{Dick:1999ed}-\cite{Cervera:2000kp}.

The transition probabilities $\nu_e \to \nu_\mu$ and $\nu_\mu \to \nu_e$
are extremely sensitive to $\theta_{13}$ and $\delta$: this is what is called 
the ``{\it golden measurement at the Neutrino Factory}''~\cite{Cervera:2000kp} 
and can be easily studied by searching for wrong-sign muons,
provided the considered detector has a good muon charge 
identification capability. The determination of ($\theta_{13},\delta$) from this 
channel is not at all free of ambiguities: in~\cite{Burguet-Castell:2001ez} 
it was shown that, for a given physical input parameter pair ($\bar \theta_{13},\bar \delta$),
measuring the oscillation probability for $\nu_e \to \nu_\mu$ and
$\bar \nu_e \to \bar \nu_\mu$ will generally result in two allowed
regions of the parameter space. The first one contains the physical
input parameter pair and the second, the ``intrinsic ambiguity'', is
located elsewhere. Worse than that, new degeneracies have later been
noticed~\cite{Minakata:2001qm,Barger:2001yr}, resulting from our
ignorance of the sign of the $\Delta m^2_{atm}$ squared mass
difference and from the approximate $[\theta_{23}, \pi/2 -
\theta_{23}]$ symmetry for the atmospheric angle. In general, for each
physical input pair the measure of $P(\nu_e \to \nu_\mu)$ and $P(\bar
\nu_e \to \bar \nu_\mu)$ will result in eight allowed regions of the
parameter space, the {\it eightfold-degeneracy}~\cite{Barger:2001yr}.

From what learned in the previous papers \cite{Minakata:2003ca}-\cite{Donini:2002rm}
we conclude that the optimal combination to deal with the eightfold-degeneracy
consists in taking advantage of all the neutrino beams
produced in a Neutrino Factory Complex (i.e. factory plus detectors).
The Neutrino Factory Complex that we consider consists of a SPL-like superbeam 
\cite{Gomez-Cadenas:2001eu} and a 50 GeV muon storage ring \cite{Apollonio:2002en}, plus
a network of three detectors of different technology: 
\begin{enumerate}
\item a 40 Kton Magnetized Iron Detector (MID) at $L = 2810$ Km, \cite{Cervera:2000vy};
\item a 4 Kton Emulsion Cloud Chamber (ECC) at $L= 732,2810$ Km, \cite{Donini:2002rm};
\item a 400 Kton Water Cherenkov (WC) at $L = 130$ Km, \cite{Blondel:2001jk}.
\end{enumerate}
This proposal, resulting from the combination of 
\cite{Burguet-Castell:2001ez,Burguet-Castell:2002qx} and \cite{Donini:2002rm}, 
corresponds to the design of a possible CERN-based Neutrino Factory Complex, 
with detectors located at the Frejus (the WC), at Gran Sasso
(the ECC) and at a third site to be defined (the MID and possibly the ECC).
Each one of these detectors is especially optimized to look for a particular signal: 
$\nu_\mu \to \nu_e$ oscillations for the 400 Kton WC, $\nu_e \to \nu_\mu$ for the 40 Kton MID
and $\nu_e \to \nu_\tau$ for the 4 Kton ECC. 

The physical parameters to be measured at the Neutrino Factory Complex
are, in the worst case (i.e. if the planned experiment are not able 
to measure some of them earlier), the two PMNS mixing matrix parameters $\bar \theta_{13}$ 
and $\bar \delta$, the sign of $\Delta_{atm}$, $\bar s_{atm}$, and the $\theta_{23}$-octant,
$\bar s_{oct}$, where
\begin{equation}
\bar s_{atm} = sign [ \Delta m^2_{atm} ] \, ; \qquad 
\bar s_{oct} = sign [ \tan (2 \theta_{23}) ] \, .
\end{equation}
Both discrete variables can assume the values $\pm 1$, depending on the physical assignments of 
the sign of $\Delta m^2_{atm}$ and of the $\theta_{23}$-octant ($s_{oct} =1$ for 
$\theta_{23} < \pi/4$ and $s_{oct} = -1$ for $\theta_{23} > \pi/4$).
The other parameters have been considered as fixed quantities, supposed to be 
known with good precision by the time when the Neutrino Factory will be operational. 
In particular: $\theta_{12} = 35^\circ$ and $\Delta m^2_\odot = 7 \times 10^{-5}$ eV$^2$; 
$\theta_{23} = 40^\circ,50^\circ $ (a generic value in the allowed 
region $\sin^2 (2 \theta_{23}) > 0.9$ with both possible octant choices) 
and $| \Delta m^2_{atm}| = 2.9 \times 10^{-3} $ eV$^2$; $A = 1.1 \times 10^{-4} $ eV$^2$/GeV. 

The experimental information consists of the number of muons in the detector with charge 
opposite to that of the muons circulating in the storage ring. 
We group the events in bins of the final muon energy $E_\mu$ 
and call $N^{g,s}(\bar \theta_{13}, \bar \delta)$ the number of ``golden'' or 
``silver'' in the i-th energy bin for the input pair ($\bar \theta_{13}, \bar \delta$) 
\cite{Cervera:2000kp}. In the case of the Superbeam, 
$N$ represents the number of electrons in the water Cherenkov, grouped in one single bin.
For a given energy bin and fixed input parameters ($\bar \theta_{13},\bar \delta$), 
we can draw a set of curves of equal number of events \cite{Donini:2002rm}
in the ($\theta_{13}, \delta$) plane, 
\begin{eqnarray}
\label{eq:ene0}
N^i_{\mu^\pm}(\bar \theta_{13}, \bar \delta; \bar s_{atm}, \bar s_{oct}) &=& 
N^i_{\mu^\pm} (\theta_{13}, \delta; s_{atm} = \bar s_{atm}, s_{oct} = \bar s_{oct}) )\, , \\
\label{eq:ene0sign}
N^i_{\mu^\pm}(\bar \theta_{13}, \bar \delta; \bar s_{atm}, \bar s_{oct} ) &=& 
N^i_{\mu^\pm} (\theta_{13}, \delta; s_{atm} = -\bar s_{atm}, s_{oct} = \bar s_{oct}) )\, , \\
\label{eq:ene0t23}
N^i_{\mu^\pm}(\bar \theta_{13}, \bar \delta; \bar s_{atm}, \bar s_{oct}) &=& 
N^i_{\mu^\pm} (\theta_{13}, \delta; s_{atm} = \bar s_{atm}, s_{oct} = -\bar s_{oct}))\, , \\
\label{eq:ene0t23sign}
N^i_{\mu^\pm}(\bar \theta_{13}, \bar \delta; \bar s_{atm}, \bar s_{oct} ) &=& 
N^i_{\mu^\pm} (\theta_{13}, \delta; s_{atm} = -\bar s_{atm}, s_{oct} = -\bar s_{oct}))\, , 
\end{eqnarray}

\begin{figure}[h!]
\begin{tabular}{cc}
\epsfxsize5cm\epsffile{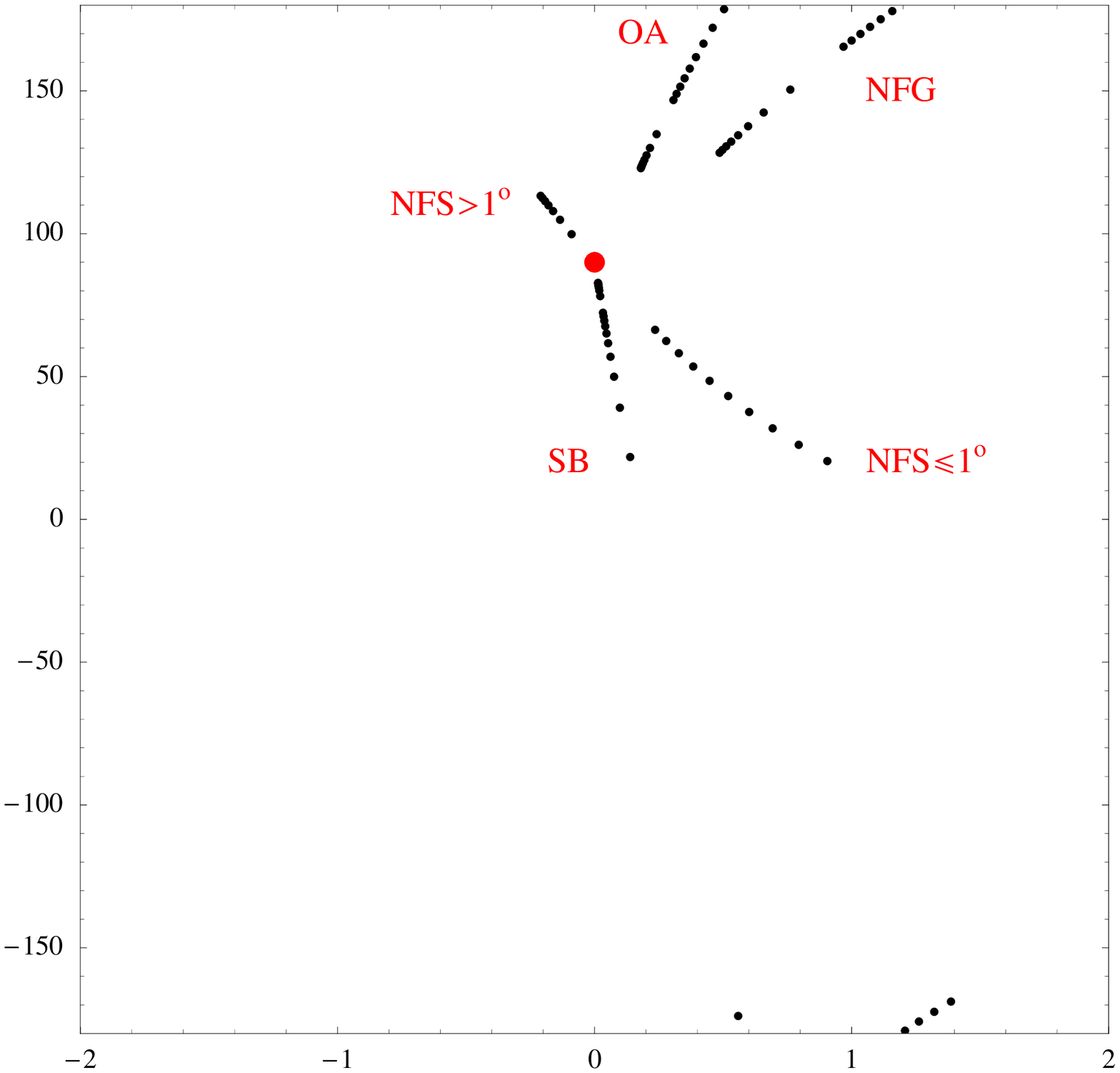} & 
\epsfxsize5cm\epsffile{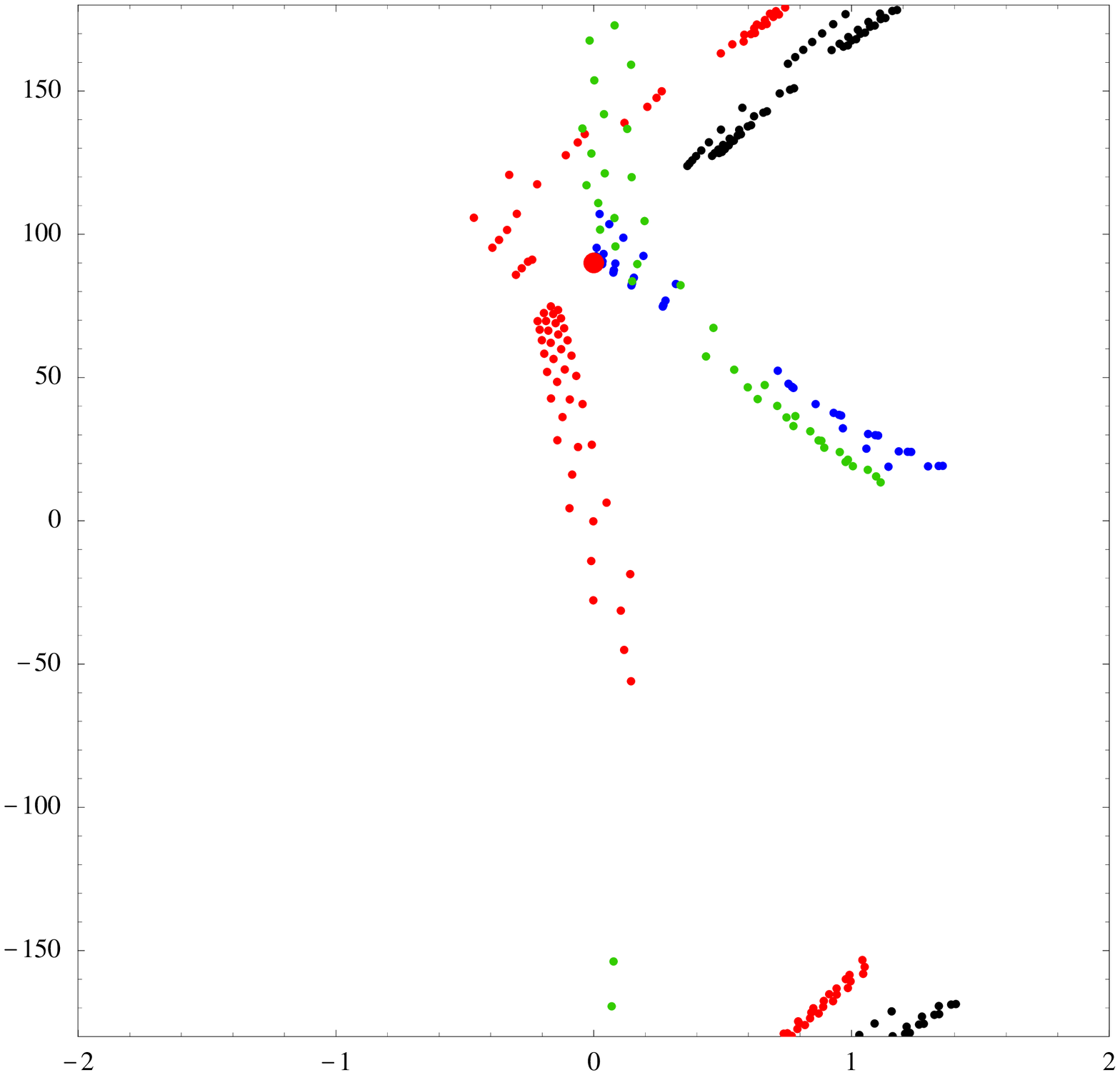}  \\
\end{tabular}
\caption{{\it 
(left) Intrinsic clone flows for the SPL and the NuMI Off-Axis
superbeams and for the NF golden and silver channels;
(right) The eightfold-degeneracy for the NF golden channel.
The thick dot is the true solution, the thin dots the clones location 
for changing $\bar \theta_{13} \in [0.1^\circ,10^\circ]$ and fixed $\bar \delta = 90^\circ$.
}}
\label{fig:clones}
\end{figure}

Following the procedure outlined in \cite{Burguet-Castell:2001ez,Donini:2002rm} we can
numerically solve eqs.~(\ref{eq:ene0})-(\ref{eq:ene0t23sign}) and found the theoretical
location of the clones in the ($\theta_{13},\delta$) plane. We present in Fig.\ref{fig:clones}
the outcome of this procedure for the different degeneracies with fixed $\bar \delta = 90^\circ$
and changing $\bar \theta_{13} \in [0.1^\circ,10^\circ]$. 
Apart from some exceptional abrupt change (remind $2 \pi$-periodicity in the $\delta$ axis), 
a small change in the input parameter $\bar \theta_{13}$ results in a small shift of the clone 
location. (Almost) continuous geometrical regions where degeneracies lie are defined 
for a given interval in $\bar \theta_{13}$, illustrating how the clones move due to a change 
in the input parameters: we will call this the ``clone flow''.
In Fig.~\ref{fig:clones}(left) we plotted the intrinsic clone flow for a set of different 
experiments and channels; in Fig.~\ref{fig:clones}(right) the clone flows for 
the eightfold-degeneracy are presented, for the NF-golden channel only. 

\begin{figure}[h!]
\begin{tabular}{cc}
\hspace{-1cm} \epsfxsize6cm\epsffile{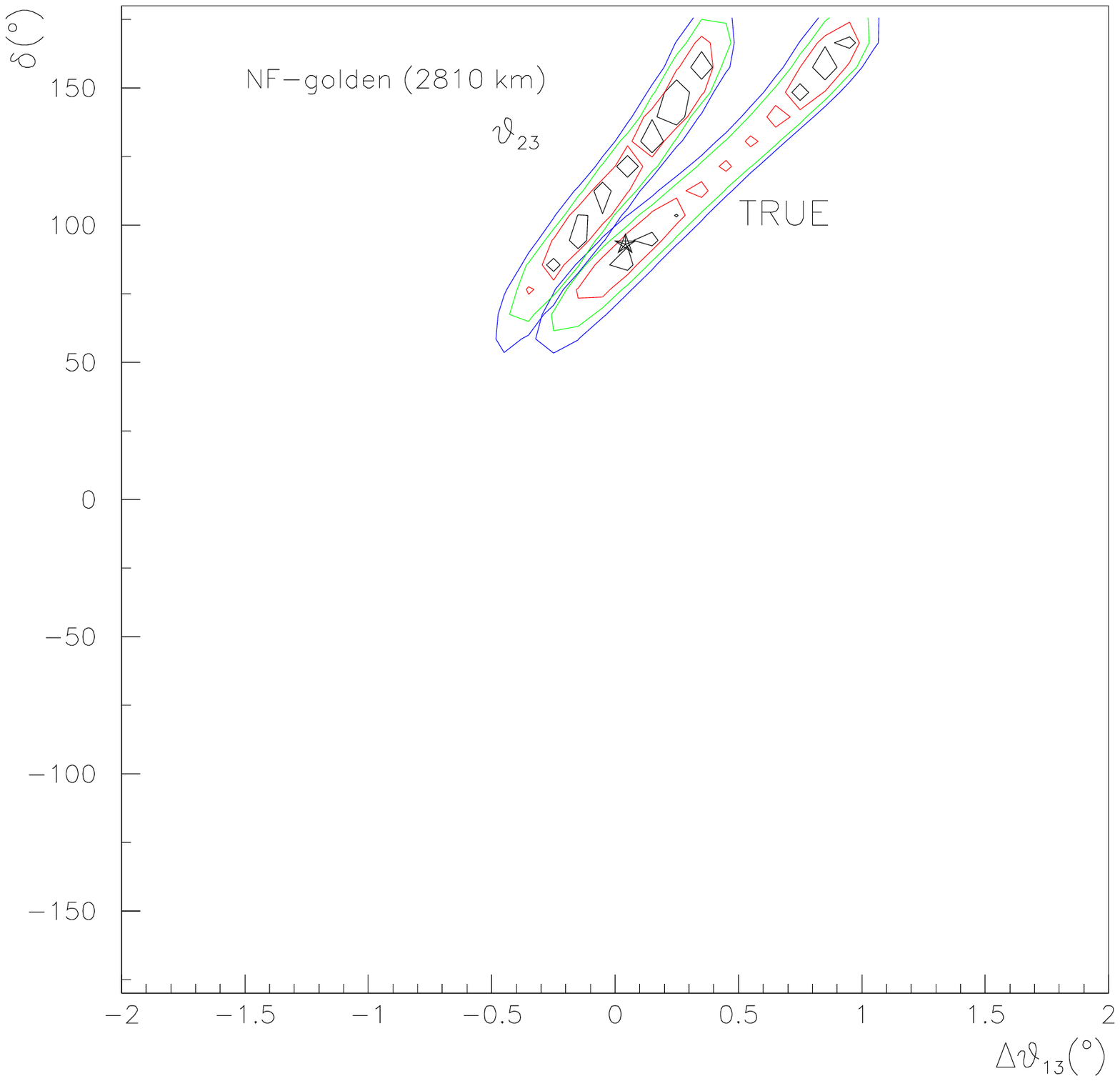} & 
              \epsfxsize6cm\epsffile{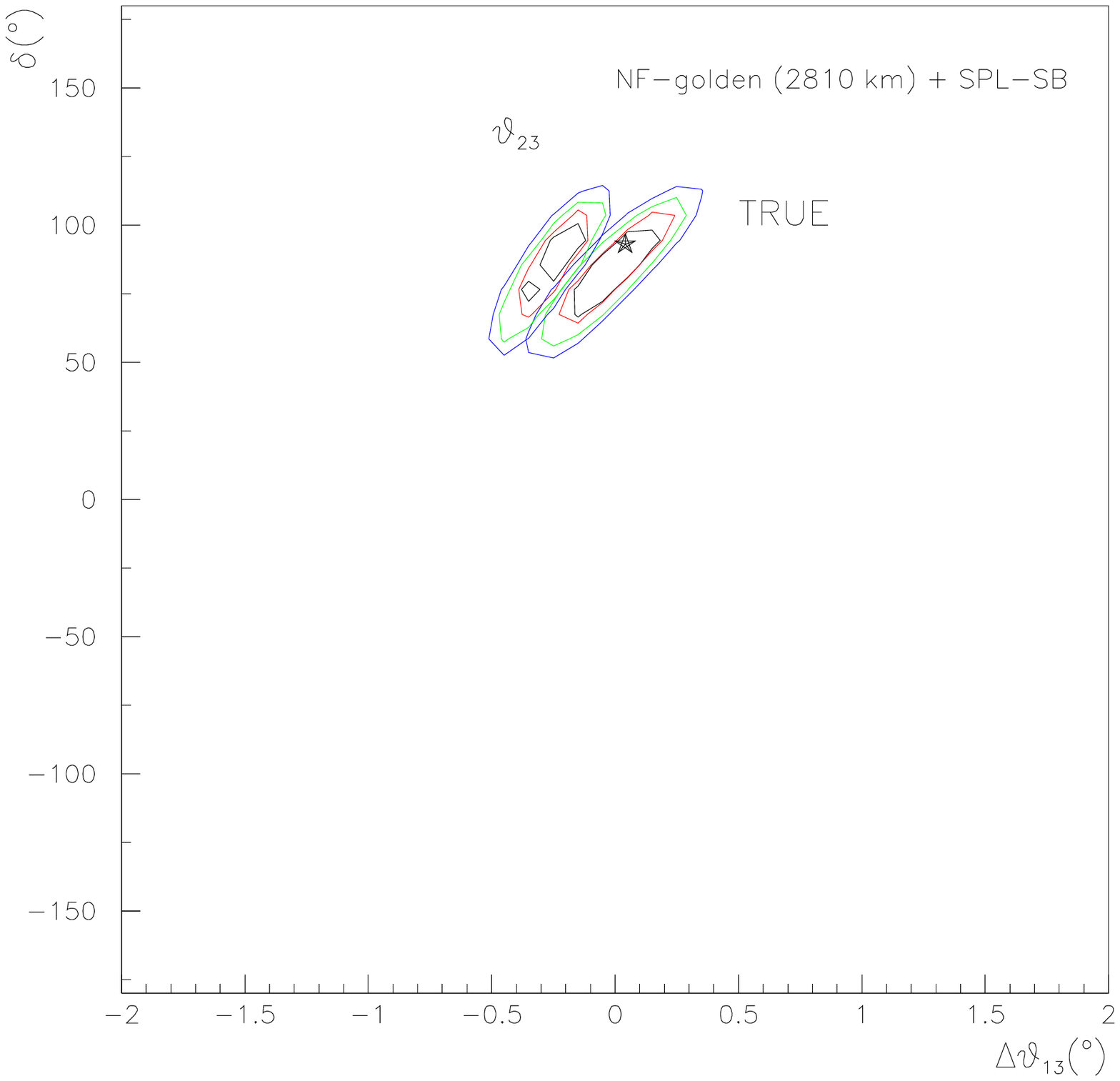} \\
\hspace{-1cm} \epsfxsize6cm\epsffile{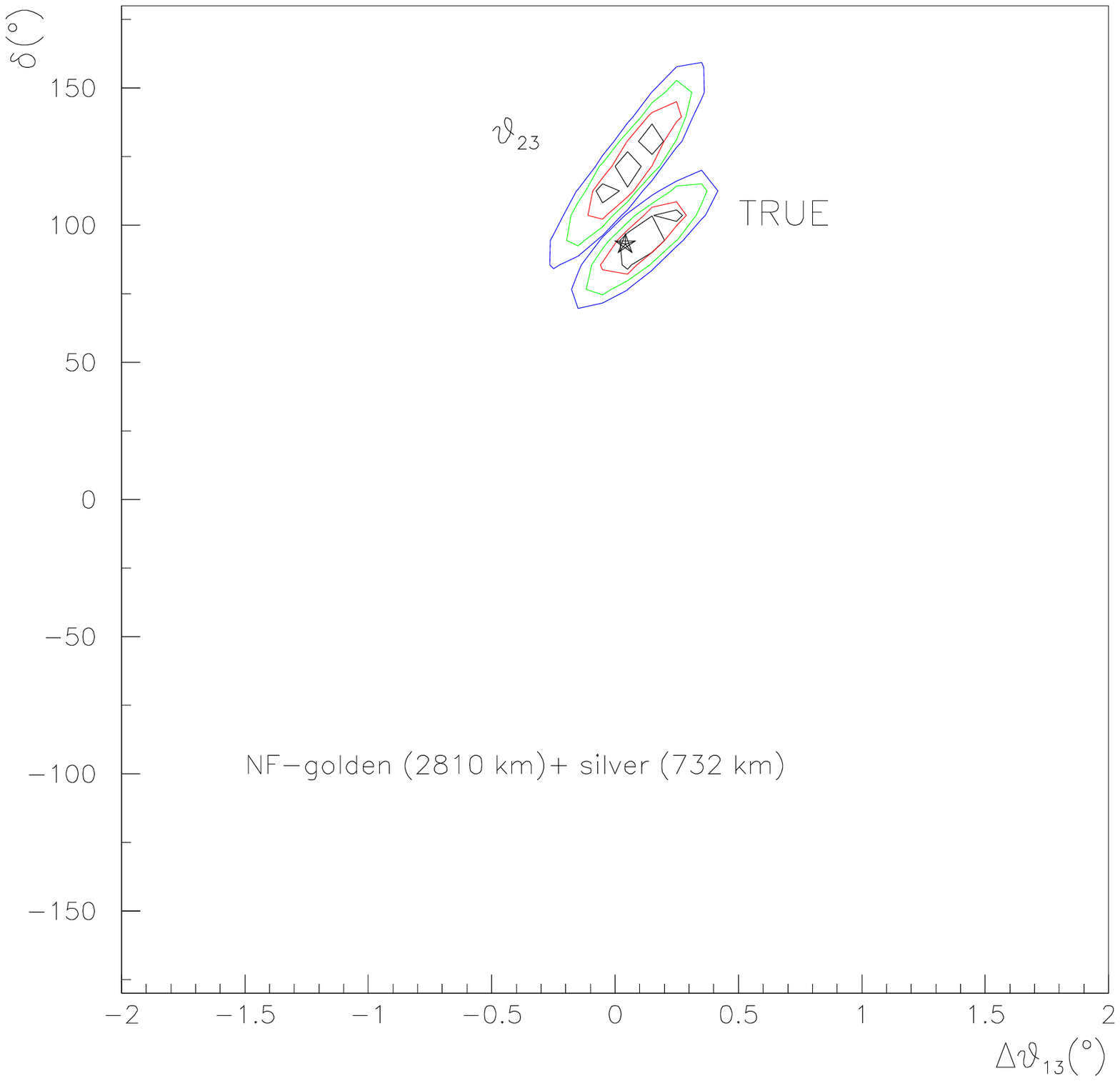} & 
              \epsfxsize6cm\epsffile{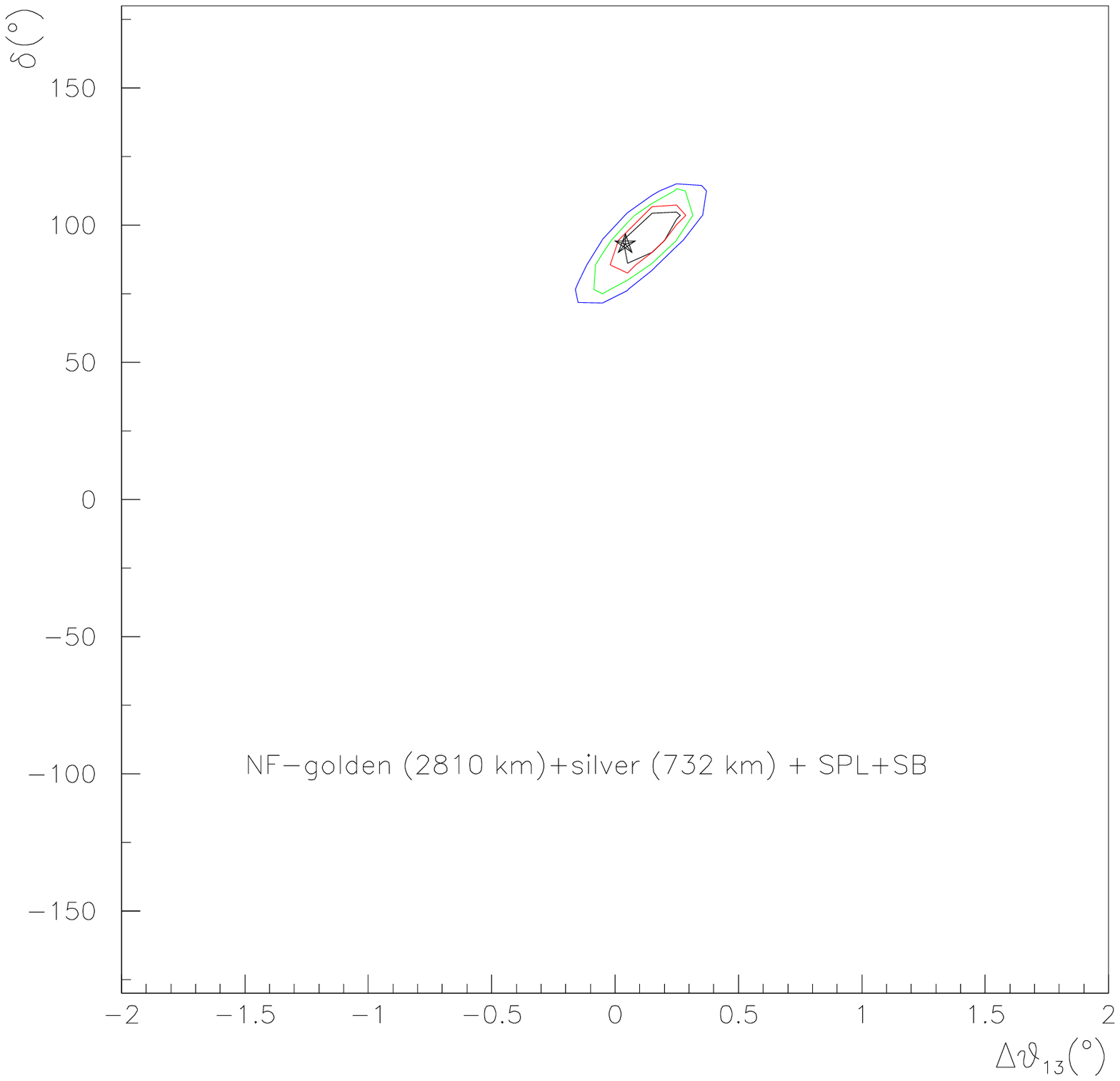} \\
\end{tabular}
\caption{\it 
The results of a $\chi^2$ fit for $\bar \theta_{13} = 2^\circ; \bar \delta = 90^\circ$.
Four different combinations of experimental data are presented: 
a) MID; b) MID plus WC; c) MID plus ECC; d) The three detectors together. 
}
\label{fig:allfits}
\end{figure}

Fig.~\ref{fig:clones}(left) shows that the combination of any two facilities solves the 
intrinsic degeneracy \cite{Burguet-Castell:2002qx,Donini:2002rm}. More difficult is
the case when all the degeneracies are treated on equal footing, Fig.\ref{fig:clones} (right),
where the need of the combination of (at least) three facilities is manifest. This is exemplified
in Fig.~\ref{fig:allfits}, where we present the outcome of combined $\chi^2$ fits performed 
as in \cite{Cervera:2000kp} for different combinations of the three detectors, 
for a fixed input pair $\bar \theta_{13} = 2^\circ, \bar \delta = 90^\circ$.
In Fig.~\ref{fig:allfits}(a) four degeneracies can be seen when using the 40 Kton MID only;
in Fig.~\ref{fig:allfits}(b) and Fig.~\ref{fig:allfits}(c) we notice how two of the 
degeneracies disappear when combining the 40 Kton MID with the 400 Kton WC or the 4 Kton ECC,
respectively; eventually, in Fig.~\ref{fig:allfits}(d) the combination of the three
detectors solve all the degeneracies reconstructing with a good precision the physical input
values. CL contours up to 4 sigma are plotted.

Two comments are in order: first, the physical input pair $\bar \theta_{13} = 2^\circ, 
\bar \delta = 90^\circ$ is generic and similar results are obtained for different input
parameters for $\bar \theta_{13} > 1^\circ$; 
second, these results, although promising, are still preliminar and
a new study where particular care is devoted to systematics in the three detectors
is currently underway \cite{superpaper}. 


\end{document}